\documentclass[11pt]{article}
\usepackage{a4wide}                      
\usepackage{epsf}                        
\usepackage{latexsym}                    
\usepackage{amsfonts}                    
\usepackage{amssymb}                     

\def\be{\begin{equation}}
\def\ee{\end{equation}}
\def\bea{\begin{eqnarray}}
\def\eea{\end{eqnarray}}

\def\part{\partial}

\def\incl{\mbox{i}}

\def\Z{\ensuremath{\mathbb{Z}}}

\def\R{\ensuremath{\mathbb{R}}}

\def\makeatletter{\catcode`\@=11}
\makeatletter
\def\mathbox#1{\hbox{$\m@th#1$}}%
\def\math@ccstyles#1#2#3#4#5#6#7{{\leavevmode
      \setbox0\mathbox{#6#7}%
      \setbox2\mathbox{#4#5}%
      \dimen@ #3%
      \baselineskip\z@\lineskiplimit#1\lineskip\z@
      \vbox{\ialign{##\crcr
             \hfil \kern #2\box2 \hfil\crcr
             \noalign{\kern\dimen@}%
             \hfil\box0\hfil\crcr}}}}
\def\mathaccstyles{\math@ccstyles\maxdimen}
\def\maththroughstyles{\math@ccstyles{-\maxdimen}}
\def\unity%
 {\maththroughstyles{.45\ht0}\z@\displaystyle {\mathchar"006C}\displaystyle 1}
\begin{document}

\rightline{FFUOV-06/10}
\rightline{FTUAM-06/09}
\rightline{hep-th/0606057}
\rightline{June 2006}
\vspace{3.5cm}

\centerline{\LARGE \bf Type II pp-wave Matrix Models from Point-like Gravitons}
\vspace{1.5truecm}

\centerline{
    {\large \bf Yolanda Lozano${}^{a,}$}\footnote{E-mail address:
                                  {\tt yolanda@string1.ciencias.uniovi.es}}
    {\bf and}
    {\large \bf Diego Rodr\'{\i}guez-G\'omez${}^{b,}$}\footnote{E-mail address:
                                  {\tt Diego.Rodriguez.Gomez@cern.ch}}
                                                   }
\vspace{.4cm}
\centerline{{\it ${}^a$Departamento de F{\'\i}sica,  Universidad de Oviedo,}}
\centerline{{\it Avda.~Calvo Sotelo 18, 33007 Oviedo, Spain}}

\begin{center}
\centerline{{\it ${}^b$Departamento de F\'{\i}sica Te\'orica C-XI, }}
\centerline{{\it Universidad Aut\'onoma de Madrid,}}
\centerline{{\it Cantoblanco, 28049 Madrid, Spain}}
\end{center}

\vspace{2.5truecm}

\centerline{\bf ABSTRACT}
\vspace{.5truecm}

\noindent
The BMN Matrix model can be regarded as a theory of
coincident M-theory gravitons, which expand by Myers dielectric effect into
the 2-sphere and 5-sphere giant graviton vacua of the theory.
In this note we show that, in the same fashion, 
Matrix String theory in Type IIA pp-wave backgrounds 
arises from the action for coincident Type IIA gravitons.
In Type IIB, we show that the action for coincident gravitons in the
maximally supersymmetric pp-wave background gives rise to a
Matrix model which supports fuzzy 3-sphere giant graviton vacua
with the right behavior in the classical limit. We discuss the relation
between our Matrix model and the Tiny Graviton Matrix theory of
hep-th/0406214.


\newpage

\section{Introduction}

Taking the point of view that the BFSS Matrix model \cite{BFSS} can be
regarded as a theory of 
coincident M-theory gravitons, we showed in \cite{JL2}  that using the action for coincident gravitons
proposed therein
it is possible to go beyond the linear order approximation of \cite{KT}. This
action was successfully used in \cite{JL2,JLR3} for the study of giant gravitons \cite{GST,GMT}
in $AdS_m\times S^n$ backgrounds,
which are not linear perturbations to Minkowski.
Moreover,  in
the M-theory maximally supersymmetric pp-wave background \cite{FP}, this action, besides 
reproducing the BMN Matrix model \cite{BMN}, predicts
a new quadrupolar coupling to the M-theory 6-form potential, which supports the so far elusive
fuzzy 5-sphere giant graviton solution \cite{LR}.

In this paper we will focus on Type II pp-wave Matrix models. These models share with the
BMN Matrix model the removal of the flat directions and the existence of a large class of
giant graviton supersymmetric vacua.
We will mainly concentrate on two backgrounds: the Type IIA background that is obtained from the maximally
supersymmetric pp-wave background of M-theory after dimensional reduction
\cite{Michelson}, and  the 
maximally supersymmetric Type IIB pp-wave background \cite{BFHP}.
We will simply refer to them as the Type IIA and Type IIB pp-wave backgrounds.

Several approaches have been taken in the literature for the study of Type II pp-wave
Matrix models.
Matrix String theory in the Type IIA pp-wave background has been studied in \cite{SY,DMS}.
The approach in \cite{SY} is to start from
the supermembrane
action in the maximally supersymmetric pp-wave background of M-theory, and then use 
the correspondence law of \cite{SY2} to reduce it to ten dimensions.
Reference \cite{DMS} constructs it, in turn, from the BMN Matrix action,
using the 9-11 flip \cite{DVV}. Matrix String models in more general pp-wave
backgrounds have also been considered in \cite{Bonelli}, by studying 
certain deformations of ten dimensional $N=1$ SYM. These models include the BMN Matrix
model when dimensionally reduced to one dimension, as well as the Matrix String theory 
of \cite{SY,DMS}, in two dimensions. In this reference a possible
deformation of the IKKT Matrix String theory \cite{IKKT} which could
be suitable for the study of  Type IIB
pp-wave backgrounds was also considered.
General features about a Matrix String theory in the maximally supersymmetric pp-wave background
of Type IIB were also discussed in \cite{Gopakumar}\footnote{See also \cite{Verlinde}.}. 
A Matrix String theory for this background was however not explicitly
constructed till reference \cite{SJ}.

The  approach taken in \cite{SJ} is to regularize the light-cone 3-brane action in the 
Type IIB pp-wave background, in close analogy to the derivation in
\cite{DSV} of the BMN Matrix model from the light-cone
supermembrane action. 
The light-cone 3-brane carries $N$ units of light-cone momentum, and some of its vacua are finite
size 3-branes with zero light-cone energy, i.e. giant
gravitons\footnote{Fixing the light-cone gauge 
corresponds to going to the rest frame of the giant graviton. There is another solution consisting on a zero-size 3-brane
with the same energy, i.e. a point-like graviton.} \cite{SS}.
In close analogy to the description in \cite{BMN}, \cite{SJ} proposes
a description of the 3-sphere  vacua in terms of $N$ expanding
gravitons, each carrying one unit of light-cone momentum,
the so-called {\it tiny} gravitons. The resulting Matrix model, a one dimensional $U(N)$ gauge theory, is referred as the Tiny Graviton
Matrix theory.

Keeping in mind
that the BMN Matrix model can be regarded as a theory of
coincident M-theory gravitons \cite{LR}, which expand by Myers dielectric effect
into the 2-sphere and 5-sphere giant graviton vacua of the theory,
one would expect that the Tiny Graviton Matrix theory of \cite{SJ} could, in the same fashion,
 be regarded as a theory of Type IIB coincident gravitons, which expand by dielectric
effect into the 3-sphere vacua. The tiny gravitons of \cite{SJ} would then
simply be coincident Type IIB point-like gravitons. In fact, it was shown in 
\cite{JLR,JLR2}
that the giant graviton solutions of the $AdS_5\times S^5$ and
$AdS_3\times S^3\times T^4$ Type IIB backgrounds can be described
microscopically using the action for coincident gravitons constructed
in \cite{JLR}. 
This action contains the right multipole
moment couplings to explain the expansion of the gravitons into 
fuzzy 3-spheres or fuzzy cylinders, respectively. 
Moreover, this action is a $U(N)$ gauge theory, in which
the non-Abelian vector field is associated to (wrapped) D3-branes
``ending'' on the gravitons\footnote{The D3-branes 
are wrapped on two isometric directions of the action, which
in the pp-wave background are the light-cone direction and a combination
of the two $S^1$ fibres
in the decomposition of the 3-spheres of the background as $U(1)$ bundles over $S^2$.}. In this
construction each graviton carries one unit of light-cone momentum, and
the enhancement from the $U(1)$ gauge theory, for a single graviton, to a $U(N)$ gauge 
theory,
for $N$ gravitons, takes place identically than in a  system of coincident D-branes. This 
is in agreement with the discussion in \cite{SJ}, and with the results in \cite{SS}. 

In this article we pursue further the line of research initiated in \cite{LR}, and
show that the Matrix models that have been constructed in the literature in 
Type II pp-wave
backgrounds can be obtained from the actions for Type II coincident
gravitons constructed in \cite{JL2,JL1,JLR}.  Matrix String theory in
the Type IIA pp-wave background \cite{SY,DMS} is reproduced exactly, whereas in
Type IIB we obtain a new Matrix model which supports fuzzy 3-sphere giant graviton
solutions with the right behavior in
the large $N$ limit. We discuss with some detail the relation between this
Matrix model and the Tiny Graviton Matrix theory of \cite{SJ} throughout the paper.

The article is organized as follows. Section 2 is devoted to the study of the type IIA pp-wave Matrix
 model. After briefly reviewing the background in subsection 2.1 we
 present the action describing Type IIA coincident 
gravitons in subsection 2.2. This action is a completion of the truncated action derived
 in \cite{JL2,JL1}, which, as we discuss, is not suitable for the study of this
 background. We particularize the action 
constructed in subsection 2.2 to the Type IIA pp-wave background in
 subsection 2.3 and show the perfect agreement with the matrix actions of
\cite{SY,DMS}. Finally in subsection 2.4 we discuss some of the fuzzy
 sphere vacuum solutions. Section 3 is devoted to the study of the Type IIB
 pp-wave Matrix model. We start in subsection 3.1 
by rewriting the background in coordinates adapted to our construction.
 In subsection 3.2 we recall the action for type IIB
 coincident gravitons of \cite{JLR}. We see that 
this action is adequate for the study of the pp-wave background. In
 subsection 3.3 we present our proposal 
for the Matrix model. Finally we summarize in subsection 3.4 some of the fuzzy 3-sphere vacua of the model. We end in section 4 with some conclusions.

\section{The Type IIA Matrix Model}

\subsection{The background}

We consider the Type IIA pp-wave background:

\begin{eqnarray}
\label{IIAback}
&&ds^2=-2dx^+ dx^- -\beta
(dx^+)^2+dx_1^2+\dots +dx_8^2\, ,\nonumber\\
&&C^{(1)}_+=-\frac13 \mu x^4\, ,\qquad C^{(3)}_{+ij}=-\frac{\mu}{3}
\epsilon_{ijk}x^k\, ; \, i,j,k=1,2,3
\end{eqnarray}

\noindent which is obtained by reducing along an isometric SO(6) direction
the maximally supersymmetric pp-wave background of M-theory
\cite{Michelson}. This background preserves 24 of the original 32
supersymmetries \cite{SS,HS}.
Here 

\begin{equation}
\beta=(\frac{\mu}{3})^2(x_1^2+\dots +x_4^2)+(\frac{\mu}{6})^2
(x_5^2+\dots +x_8^2)\, .
\end{equation}

We are interested in constructing a Matrix model describing the
dynamics of gravitational waves in this background, in the sector
with momentum $p^+=-p_-=N/R$.

\subsection{The action for Type IIA gravitons}

An action describing coincident gravitational waves in Type IIA backgrounds has
been constructed to linear order in the background fields  in \cite{JL1}.  
This action is, however, not suitable for the study of the Type IIA background
(\ref{IIAback}), because $(C^{(1)}_+)^2$ contributes with a quadratic power of
the $x^4$ transverse scalar and therefore gives a contribution to the leading order
expansion of the action (see next subsection).
In this section we construct an action for Type IIA waves that goes
beyond the linear approximation, and is suitable for
the study of the Type IIA pp-wave background (\ref{IIAback}) to the
desired order.

Our starting point is the action for coincident
M-theory gravitational waves 
constructed in \cite{JL2}. This action goes beyond the linear order
approximation, and has the same regime of validity of Myers action
for coincident D-branes \cite{Myers}\footnote{We refer the reader to reference
\cite{JL2} for more details.}:

\begin{eqnarray}
\label{Mppwave}
S&=&-\int d\tau\, {\rm STr}\Bigl\{k^{-1}
\sqrt{-\Bigl[g_{\mu\nu}{\cal D}_\tau X^\mu {\cal D}_\tau X^\nu
+E_{\tau i}(Q^{-1}-\delta)^i_k E^{kj}E_{j\tau}\Bigr]{\rm det}Q}
\Bigr\}+\nonumber\\
&&+\int d\tau\, {\rm STr}\Bigl\{P\Bigl[k^{-2}k^{(1)}\Bigr]-iP\Bigl[(\incl_X
\incl_X)C^{(3)}\Bigr]+\frac12 P\Bigl[(\incl_X \incl_X)^2\incl_k C^{(6)}\Bigr]
+\dots\Bigr\}
\end{eqnarray}

\noindent where

\begin{equation}
E_{\mu\nu}=g_{\mu\nu}+k^{-1} (\incl_k C^{(3)})_{\mu\nu}\, , \qquad
Q^i_j=\delta^i_j+ik [X^i,X^k]E_{kj}\, , \qquad i=1,\dots 9\, .
\end{equation}

\noindent Here we have taken units in which the
tension of a single graviton is equal to one.
In this action the direction of
propagation of the waves appears as a special isometric direction,
with Killing vector $k^\mu$. $k^2$ and $k^{(1)}$ are defined as
$k^2=g_{\mu\nu}k^\mu k^\nu$ and $k^{(1)}_\mu=g_{\mu\nu}k^\nu$.
$k^{-2}k^{(1)}$ is then the momentum operator along the isometric
direction.
Consistently with this isometry the pull-backs into the worldvolume
are taken with gauge covariant derivatives \cite{BJO}

\begin{equation}
\label{covder}
{\cal D}_\tau X^\mu=\partial_\tau X^\mu -k^{-2}k_\nu 
\partial_\tau X^\nu k^\mu\, .
\end{equation}

\noindent In this way the dependence
on the isometric direction is effectively eliminated from the action.
This action is in fact a gauge fixed action in which
the $U(N)$ vector field, associated to M2-branes (wrapped on
the direction of propagation) ending on the waves, has been taken to
vanish, $A_\tau=0$. In this gauge 
$U(N)$ covariant derivatives reduce to ordinary derivatives, and 
gauge covariant derivatives can be defined using ordinary derivatives as in
(\ref{covder}).

We can obtain the action describing coincident Type IIA gravitons by reducing
the action (\ref{Mppwave}) along a transverse direction.
Applying the dimensional reduction rules to the
eleven dimensional metric it is clear that quadratic
powers of the RR 1-form potential show up.
Given that in the pp-wave background $C^{(1)}_+=-\mu x^4/3$, these couplings
have to be included in order to construct a Matrix model which contains
quadratic powers of the transverse scalars.
For simplicity we restrict ourselves to backgrounds in which 
the NS-NS 2-form potential vanishes, all components but the time
component and the component on the direction of propagation of the RR
1-form potential are zero, and $k_i=0$.
This is indeed a suitable truncation for the pp-wave background
(\ref{IIAback}).

We obtain for the BI action

\begin{eqnarray}
S^{\rm BI}&=&-\int d\tau\, {\rm STr} \Bigl\{ \frac{1}{\sqrt{k^2+e^{2\phi}(\incl_k C^{(1)})^2}}
\sqrt{\Bigl(\unity - (k^2+e^{2\phi}(\incl_k
  C^{(1)})^2)[A,X]^2\Bigr)
{\rm det}Q}   \nonumber\\
&& .\sqrt{-\Bigl[g_{\mu\nu}{\cal D}_\tau X^\mu {\cal D}_\tau X^\nu+\frac{k^2
      e^{2\phi}}{k^2+e^{2\phi}(\incl_k C^{(1)})^2}\Bigl(C^{(1)}_\mu {\cal
      D}_\tau X^\mu + F\Bigr)^2\Bigr]}\Bigr\}
\label{BIIIA}      
\end{eqnarray}

\noindent Here
 
\begin{eqnarray}
&&E_{\mu\nu}=g_{\mu\nu}+\frac{e^{\phi}}{\sqrt{k^2+e^{2\phi}(\incl_k C^{(1)})^2}}
(\incl_k C^{(3)})_{\mu\nu}\nonumber\\
&&Q^i_j=\delta^i_j + i[X^i,X^k]e^{-\phi}\sqrt{k^2+e^{2\phi}(\incl_k
    C^{(1)})^2} E_{kj}
\end{eqnarray}

\noindent and $i$ now runs from 1 to 8.
$A$ is the scalar field that comes from the reduction of the eleventh transverse
direction and $F$ is its field strength. $F$ forms an invariant field strength
with the pull-back of the RR 1-form potential, and therefore $A$ is associated to 
D0-branes ending on the waves.
The first square root in (\ref{BIIIA}) comes from the reduction of the determinant of
the nine dimensional $Q$ matrix, whereas the second square root comes from the
reduction of the pull-back of the metric. We should mention that in this action we have made a
further truncation. We have omitted those
terms coming from the reduction of $E_{\tau i}(Q^{-1}-\delta)^i_k
E^{kj}E_{j\tau}$. This is justified because these terms contribute to higher order on the transverse
scalars.

Dimensionally reducing the CS action we get:

\begin{eqnarray}
\label{CSIIA}
S^{\rm CS}&=&\int d\tau {\rm STr}\Bigl\{P\Bigl[k^{-2} k^{(1)}\Bigr]+
\frac{e^{2\phi} \incl_k C^{(1)}}{k^2+e^{2\phi}(\incl_k C^{(1)})^2}
\Bigl(P\Bigl[C^{(1)}\Bigr] + F\Bigr)-iP\Bigl[(\incl_X \incl_X)C^{(3)}\Bigr]
\nonumber\\
&&+\frac12 
P\Bigl[(\incl_X \incl_X)^2 \incl_k B^{(6)}\Bigr]+\dots\Bigr\}
\end{eqnarray}

\noindent {}From here we see that the Type IIA gravitons propagate along the same
isometric direction, consistently with the
fact that the isometry is
inherited when we reduce along a transverse direction.
We also find dielectric couplings to the RR 3-form and the NS-NS 6-form
potentials, which would be responsible for the expansion of the gravitons into
D2-branes and NS5-branes 
in suitable backgrounds. We have omitted couplings to
higher order background potentials and products of different background
fields contracted with the non-Abelian scalars  because they will not play a role
in the pp-wave background that we consider in this paper.

\subsection{The Matrix model}

We can now 
particularize the actions (\ref{BIIIA}) and (\ref{CSIIA}) to the 
Type IIA pp-wave background (\ref{IIAback}). We are interested in describing
waves with non-vanishing light-cone momentum $p_-$. However the
actions (\ref{BIIIA}) and (\ref{CSIIA}) are singular for the choice
$k^\mu=\delta^\mu_-$, since $k^2=g_{--}$ vanishes in the pp-wave
background. A natural way to regularize the action is to undo the
Penrose limit, keep the waves propagating in the $\psi$-direction of
the original eleven dimensional AdS background, and finally take the Penrose limit
$L\rightarrow\infty$.  Since, in the notation of (\ref{IIAback}), the light-cone
coordinates are related to the AdS time and $\psi$ coordinate through

\begin{equation}
x^+=\frac{3}{2\mu}(t+\psi)\, ,\qquad x^-=\frac{\mu L^2}{3} (t-\psi)\, ,
\end{equation}

\noindent $p_-$ and $p_\psi$ are related through 

\begin{equation}
p_\psi=-p_- \frac{\mu L^2}{3}\, .
\end{equation}

\noindent If we take
the gravitons propagating along the $\psi$ direction with momentum 
$p_\psi=N$ we have that $p_-=-N/R$ if $R$ is related to the AdS
radius and the mass scale as $R=\mu L^2/3$.

Therefore, we take $k^\mu=\delta^\mu_\psi$ in the actions (\ref{BIIIA}) and
(\ref{CSIIA}). Moreover, we take light-cone gauge and identify $x^+$ with
the worldline time. Expanding the action (\ref{BIIIA}) to leading order (in
$\lambda=2\pi \alpha^\prime$) we find\footnote{Note that in our units
$\lambda=1$. This approximation is however the usual one taken in
non-Abelian BI actions (see \cite{Myers}), based on the fact that these
actions are good to describe the system of branes when they are distances away
less than the Planck length.}

 \begin{eqnarray}
S^{\rm BI}&=&-\frac{\mu}{3}\int dx^+ {\rm STr}\Bigl\{\unity
-\frac{L^2}{4}[X,X]^2
-\frac{L^2}{2}[A,X]^2+\frac{i}{2}\epsilon_{ijk}X^iX^jX^k+
\frac{9{\tilde \beta}^2}{4\mu^2L^2}-\frac{9}{2\mu^2L^2}\dot{X}^2
\nonumber\\
&&-\frac{9}{2\mu^2L^2}F\Bigl(F-\frac{\mu}{3}X^4\Bigr)\Bigr\}
\end{eqnarray}

\noindent and

\begin{equation}
S^{\rm CS}=\int dx^+{\rm STr}\Bigl\{\frac{\mu}{3}\Bigl(1-\frac{9{\tilde
    \beta}}{4\mu^2 L^2}\Bigr)-\frac{1}{2L^2}X^4 F-
i\frac{\mu}{6}\epsilon_{ijk}X^iX^jX^k\Bigr\}
\end{equation}

\noindent where
 
\begin{equation}
{\tilde \beta}=\beta-\frac{\mu^2}{9}(X^4)^2\, ,
\end{equation}

\noindent $i=1,2,3$ and we denote the non-Abelian transverse scalars with capital letters.

Therefore, the final action reads

\begin{equation}
\label{IIAmatrix}
S=\int dx^+ {\rm STr} \Bigl( \frac{1}{2R} \dot{X}^2-\frac{{\tilde
    \beta}}{2R}
+\frac{R}{4}[X,X]^2+\frac{R}{2}[A,X]^2+\frac{1}{2R}
    F\Bigl(F-\frac{2\mu}{3}X^4\Bigr)
-i\frac{\mu}{3}\epsilon_{ijk}X^iX^jX^k\Bigr)
\end{equation}

\noindent where we have substituted $R=\mu L^2/3$. This action is in
perfect agreement with the results in \cite{SY} and \cite{DMS} when
one makes the truncation $\partial_\sigma X^\mu=0$ (see the discussion
below).

Indeed, it was shown in \cite{BJL,JL1}
that Matrix String theory has an
alternative interpretation as describing the dynamics of coincident
Type IIA gravitons. 
The idea in \cite{BJL,JL1} is that since Matrix String theory describes
string states with fixed light cone momentum, it could, in some limit,
effectively describe gravitons.
Explicitly, Matrix String theory is constructed by compactifying
M-theory on the 9th direction, and then performing the 9-11
flip \cite{DVV}. However, when one considers weakly curved backgrounds the
9th direction appears as a special isometric direction on which neither
the background fields nor the currents depend. This is translated into
a reduced, $SO(8)$ transverse rotationally invariant action. One can however
rewrite the action in terms of ten dimensional pull-backs into a one 
dimensional worldvolume by using the techniques of gauged sigma
models. 

The 9th direction is interpreted as the spatial worldsheet direction of
the string, $\sigma$. If one makes the truncation $\partial_\sigma X^\mu=0$
and let $k^\mu$ be the Killing vector pointing along the 9th direction
one can achieve invariance under the local isometric transformations
generated by $k^\mu$ 
by introducing gauge covariant derivatives as in (\ref{covder}).
Using gauge covariant pull-backs, constructed with these gauge
covariant derivatives, it is possible to eliminate the pull-back of
the isometric coordinate, and to reproduce the isometric
couplings in the Matrix String action in a manifestly covariant
way (see \cite{JL1}).

Consistently with this discussion we reproduce, using the action for Type IIA gravitons,
the subsector of the Matrix String theory constructed in \cite{SY,DMS} 
satisfying $\partial_\sigma X^\mu=0$.

\subsection{The fuzzy sphere solutions}

The Matrix model (\ref{IIAmatrix}) admits fuzzy 2-sphere giant graviton solutions.
One is a static fuzzy sphere located at $x^4=\dots =x^8=0$, 
identical to the 2-sphere solution
of the BMN Matrix model but from the fact that it preserves just eight
supersymmetries due to the toroidal compactification from M-theory \cite{SY}.
A more general solution is considered in \cite{DMS} in which the static fuzzy sphere
is located in an arbitrary point in the $x^4$ direction,  which preserves
as well eight supersymmetries.  In this section we are going to show however that
the $x^4$ location and
the radius of the fuzzy 2-sphere must satisfy the relation

\begin{equation}
x^4 r=-\frac{m\sqrt{N^2-1}}{2N}\, ,
\end{equation}

\noindent where $m$ is an integer, in agreement with the results found in \cite{CS} in the large $N$ limit.

Let us first discuss the solutions in \cite{CS}. In
this reference spherical D2-brane giant graviton solutions are studied using a test
D2-brane in the pp-wave background, carrying light-cone momentum.
It is found that 2-sphere solutions with non-vanishing $x^4$ are
possible when a magnetic field inducing D0-brane charge in the configuration
is switched on. Then the radius of the spherical D2-brane and its $x^4$ location  must satisfy
the relation

\begin{equation}
\label{rel}
 x^4 r=-m/2\, ,
 \end{equation}

\noindent  where $m$ is the D0-brane charge induced in the worldvolume.
 This result is supported by a microscopical calculation in terms of non-Abelian $m$
 D0-branes expanding into the fuzzy 2-sphere by dielectric effect. 
 
 We now show that a similar condition for the radius of the spherical fuzzy
 D2-brane and its $x^4$ location is predicted within the Matrix model
 description.
 
Let us start by taking in (\ref{IIAmatrix}) the fuzzy 2-sphere ansatz:

\begin{equation}
X^i=\frac{r}{\sqrt{C_N}} J^i\, , \qquad i=1,2,3
\end{equation}

\noindent with $J^i$ the generators of $SU(2)$ in an $N$ dimensional representation 
(in our notation $[J^i,J^j]=2i\epsilon^{ijk} J^k$) and
$C_N$ the quadratic Casimir in this representation.
 Take as well $X^4=$ constant and Abelian (we will denote
it as $x^4$), 
$X^5=\dots =X^8=0$, and $F$ Abelian.
We then obtain an action

\begin{equation}
S=-\frac{N}{2R}\int dx^+ \, \Bigl[ \Bigl( \frac{\mu}{3}-\frac{2Rr}{\sqrt{N^2-1}}\Bigr)
r^2-F\Bigl( F-\frac23 \mu x^4\Bigr) \Bigr]\, ,
\end{equation}

\noindent and, Legendre transforming $F$,  a Hamiltonian

\begin{equation}
\label{Hmatrix}
H=\frac{N}{2R}\Bigl[ \Bigl(\frac{\mu}{3}-\frac{2Rr}{\sqrt{N^2-1}}
\Bigr)^2r^2+\Bigl( \frac{\mu x^4}{3}+\frac{p_{A}R}{N}\Bigr)^2
\Bigr]\, ,
\end{equation}

\noindent where $p_{A}$ is the conjugate momentum of the
scalar field $A$. Recalling that $A$ has its origin on the eleventh
transverse scalar, $p_A$ is the momentum along the 
eleventh direction, which is interpreted in ten dimensions as D0-brane charge.

Now, minimizing with respect to $r$ and $x^4$, we find two zero energy solutions,
one with zero radius, the point-like graviton, and a second one with 

\begin{equation}
\label{radius}
r=\frac{\mu\sqrt{N^2-1}}{6R}\, ,
\end{equation}

\noindent which corresponds to the giant graviton. Both solutions are located at

\begin{equation}
\label{x_4value}
x^4=-\frac{3mR}{\mu N} \, ,
\end{equation}

\noindent for $m$ D0-brane charge.

Therefore, microscopically, the radius and the position in $x^4$ of the
giant graviton must be related through

\begin{equation}
x^4 r=-\frac{m\sqrt{N^2-1}}{2N}\, .
\end{equation}

\noindent This reproduces the relation (\ref{rel})  of \cite{CS} in the large $N$ limit.
Moreover, the Hamiltonian (\ref{Hmatrix}), derived using the Matrix model, coincides in
the large $N$ limit with the Hamiltonian describing the spherical D2-brane
with momentum $N$ and D0-brane charge $m$ of \cite{CS}, as we now show.

A classical spherical D2-brane in the pp-wave background (\ref{IIAback}) with $x^-=x^-(x^+)$ sitting at a
constant $x^4$ at $x^5=\dots =x^8=0$, and carrying D0-brane charge $m$, is
described by a Lagrangian

\begin{equation}
S=-4\pi T_2 \int dx^+\ \Big\{\sqrt{\frac{\mu^2}{9}\Bigl(r^2+(x^4)^2\Bigr)+2\dot{x}^-}
\sqrt{r^4+\frac{m^2}{4}}-\frac{\mu}{3}\Bigl(r^3-\frac{m}{2}x^4\Bigr)\Big\}\ ,
\end{equation}

\noindent where we have substituted

\begin{equation}
F_{\theta\phi}=\frac{m}{2}\sin{\theta}\, ,
\end{equation}

\noindent which is the magnetic field inducing $m$ D0-brane charge in the worldvolume.

Legendre transforming with respect to $\dot{x}^-$ we arrive at the following
Hamiltonian in terms of the canonically conjugated momentum $p_-$

\begin{equation}
H=-\frac{p_-}{2}\Bigl[\Bigl(\frac{\mu}{3}+\frac{4\pi
    T_2 r}{p_-}\Bigr)^2 r^2
+\Bigl(\frac{\mu x^4}{3} -\frac{2\pi T_2 m}{p_-}\Bigr)^2\Bigr]\, .
\end{equation}

\noindent This expression coincides exactly in the large $N$ limit
with the microscopical Hamiltonian
(\ref{Hmatrix}), when we take into account that in our units $2\pi T_2=1$.

 The minimum energy, $E=0$, is reached when:

\begin{equation}
r=-\frac{\mu p_-}{12\pi T_2}\, ,\qquad
x^4=\frac{6\pi T_2 m}{\mu p_-}\, ,
\end{equation}

\noindent which also agree exactly in the large $N$ limits with expressions
(\ref{radius}) and (\ref{x_4value}).

We can conclude that the classical 2-sphere solution found in \cite{CS} is correctly 
reproduced within the Matrix model description.
Since this
2-sphere giant graviton solution carries both momentum
charge (in our notation, $N$), and
D0-brane charge (in our notation, $m$) it is possible to describe it microscopically either
as $N$ gravitational waves with D0-brane charge $m$, expanding due to their (dielectric)
coupling to the RR 3-form potential, or as $m$ D0-branes moving along the $x^-$ direction
with momentum $N$ (expanding as well due to their coupling to the 3-form potential).
This explains why the microscopical
description of \cite{CS} in terms of D0-branes agrees with the Matrix model description.
The Matrix model provides however
a unified set-up for the microscopical study of giant graviton configurations in both Type
II and M theories.

 \section{The Type IIB Matrix Model}
 
 In this section we propose a Matrix model for the maximally supersymmetric pp-wave
 background of Type IIB. This Matrix model is constructed from the action describing
 coincident Type IIB gravitons.
 
 \subsection{The background}
 
 We start by recalling the form of the maximally supersymmetric pp-wave background of
 Type IIB \cite{BFHP}. 
 It arises as the Penrose limit of the $AdS_5\times S^5$ background
 
 \begin{eqnarray}
 &&ds^2=L^2(-\cosh^2{\rho}\,d\tau^2+d\rho^2+\sinh^2{\rho}\,d\Omega_3^2+d\theta^2+
 \cos^2{\theta}d\psi^2+\sin^2{\theta}d{\tilde \Omega}_3^2)\nonumber\\
 &&C^{(4)}_{\tau\alpha_1\alpha_2\alpha_3}=-L^4\sinh^4{\rho}\sqrt{g_\alpha}\, ,\qquad
 C^{(4)}_{\psi\gamma_1\gamma_2\gamma_3}=-L^4\sin^4{\theta}\sqrt{g_\gamma}
 \end{eqnarray}
 
 \noindent where $\{\alpha_i\}$ and $\{\gamma_i\}$ are, respectively,
  the angles parametrizing the 3-spheres
 contained in the $AdS$ and $S$ parts of the geometry:
 
 \begin{eqnarray}
 &&d\Omega_3^2=d\alpha_1^2+\sin^2{\alpha_1}(d\alpha_2^2+\sin^2{\alpha_2}d\alpha_3^2)\, ,
 \nonumber\\
 &&d{\tilde \Omega}_3^2=d\gamma_1^2+\sin^2{\gamma_1}(d\gamma_2^2+
 \sin^2{\gamma_2}d\gamma_3^2)\, ,
 \end{eqnarray}
 
 \noindent and $\sqrt{g}$ is the volume element on the unit 3-sphere.
 
  Defining
  
 \begin{equation}
 \label{coord}
 x^+=\frac{1}{2\mu}(\tau+\psi)\, , \qquad x^-=\mu L^2(\tau-\psi)\, , \qquad
 \rho=\frac{r}{L}\, , \qquad \theta=\frac{y}{L}
 \end{equation}
 
 \noindent and taking $L\rightarrow\infty$ one gets \cite{BFHP2}
 
 \begin{eqnarray}
\label{IIBppwave0}
 &&ds^2=-2dx^+ dx^- - \mu^2(r^2+y^2)(dx^+)^2+dr^2+r^2 d\Omega_3^2+dy^2+
 y^2 d{\tilde \Omega}_3^2 \nonumber\\
 &&C^{(4)}_{+\alpha_1\alpha_2\alpha_3}=-\mu r^4\sqrt{g_\alpha}\, ,\qquad
 C^{(4)}_{+\gamma_1\gamma_2\gamma_3}=-\mu y^4\sqrt{g_\gamma}\, .
 \end{eqnarray}

 We are interested in constructing a Matrix model describing the dynamics of gravitons
in this background in the sector with momentum $p^+=-p_-=N/R$. 
In order to do this it is convenient to describe the 3-spheres 
in (\ref{IIBppwave0}) as Hopf-fiberings, $p:S^3\rightarrow S^2$, using the round 
 metric for $S^3$:
 
 \begin{equation}
 d\Omega_3^2=\frac14 \Bigl((d\chi-A)^2+d\Omega_2^2\Bigr)
 \end{equation}
 
 \noindent where, in Euler angles:
 
 \begin{equation}
 A=-\cos{\chi_1}d\chi_2\, , \qquad d\Omega_2^2=d\chi_1^2+\sin^2{\chi_1}
 d\chi_2^2\, .
 \end{equation}
 
\noindent Using now Cartesian coordinates to describe the 
 2-spheres we can write
 the background metric and potentials (\ref{IIBppwave0}) as
 
 \begin{eqnarray}
 \label{IIBppwave}
 &&ds^2=-2dx^+dx^--\mu^2(r^2+y^2)(dx^+)^2+dr^2+\frac{r^2}{4}
 \Bigl((d\chi-A)^2+dx_1^2+dx_2^2+dx_3^2\Bigr)+\nonumber\\
 &&+dy^2+\frac{y^2}{4}
 \Bigl((d{\tilde \chi}-{\tilde A})^2+dz_1^2+dz_2^2+dz_3^2\Bigr)\, ,
 \nonumber\\
 &&C^{(4)}_{\chi + ij}=\frac18 \mu r^4\epsilon_{ijk}x^k\, , \qquad i,j=1,2,3\, ,
 \nonumber\\
 &&C^{(4)}_{{\tilde \chi}+ab}=\frac18 \mu y^4\epsilon_{abc}z^c\, ,
  \qquad a,b=1,2,3
 \end{eqnarray}
 
 \noindent where $\vec{x}$ and $\vec{z}$ parametrize points in 
 $\R^3$. In these
 coordinates:
 
 \begin{eqnarray}
 &&A= -\frac{x_3}{x_1^2+x_2^2}(x_1dx_2-x_2dx_1)\nonumber\\
 &&{\tilde A}=-\frac{z_3}{z_1^2+z_2^2}(z_1dz_2-z_2dz_1)\, .
 \end{eqnarray}
 
Note that in these coordinates we have reduced the explicit invariance
of the background from $U(1)^2\times (SO(4))^2$ to
$U(1)^2\times (SO(3)\times U(1))^2$, though the
whole invariance should still be present in a non-manifest way.

 \subsection{The action for Type IIB gravitons}
 
 The action describing coincident Type IIB gravitons was constructed in \cite{JLR}.
 Like the action for Type IIA gravitons, it goes beyond the linear order approximation
 and has the same regime of validity than Myers action for coincident D-branes. 
 It is given by:
 
 \begin{eqnarray}
\label{IIBaction}
S&=&-\int d\tau\, {\rm STr}\Bigl\{k^{-1}
\sqrt{-\Bigl[g_{\mu\nu}{\cal D}_\tau X^\mu {\cal D}_\tau X^\nu
+E_{\tau i}(Q^{-1}-\delta)^i_k E^{kj}E_{j\tau}\Bigr]{\rm det}Q}
\Bigr\}+\nonumber\\
&&+\int d\tau\, {\rm STr}\Bigl\{P\Bigl[k^{-2}k^{(1)}\Bigr]-iP\Bigl[(\incl_X
\incl_X)i_l C^{(4)}\Bigr]
+\dots\Bigr\}
\end{eqnarray}
 
 \noindent where
 
 \begin{equation}
 \label{IIBaction2}
 E_{\mu\nu}=g_{\mu\nu}-k^{-1}l^{-1}e^{\phi}
 (\incl_k \incl_l C^{(4)})_{\mu\nu}\,, \qquad
 Q^i_j=\delta^i_j+i[X^i,X^k]e^{-\phi} kl 
E_{kj}\, , \qquad i=1,\dots 7
 \end{equation}
 
 \noindent and the gauge covariant derivatives are defined as
 
 \begin{equation}
\label{covder2}
 {\cal D}_\tau X^\mu=\partial_\tau X^\mu-k^{-2} k_\nu \partial_\tau X^\nu k^\mu
 -l^{-2} l_\nu \partial_\tau X^\nu l^\mu\, .
 \end{equation}
 
 \noindent A detailed discussion of this action can be found in \cite{JLR}. Like the action for Type IIA
 gravitons, the direction of propagation appears as a special isometric direction, with
 Killing vector $k^\mu$. In this case however there is a second isometric direction, with
 Killing vector $l^\mu$, which is inherited from the T-duality transformation involved in the
 construction. Although in the Abelian limit the dependence in this direction can be restored,
 this does not happen in the non-Abelian case (see the discussion in \cite{JLR}).
 Note that it is precisely due to the existence of this second isometry that the RR 4-form potential can couple in the action (\ref{IIBaction}). Otherwise the dielectric couplings in (\ref{IIBaction}) and in (\ref{IIBaction2}) would not be possible\footnote{One could expect in principle a coupling of the form $(i_X i_X) i_k C^{(4)}$ in the CS action. However, such a coupling does not arise in the T-duality transformation involved in the construction of (\ref{IIBaction}), and, moreover, it vanishes in the $AdS_5\times S^5$ background. Indeed, in this background, the coupling that is responsible for the existence of the dual giant graviton solution is $(i_X i_X) i_l C^{(4)}$ \cite{JLR}.}.
 Therefore, the action (\ref{IIBaction}) is suitable for the study
 of gravitons which propagate in backgrounds with isometric directions. This is indeed the case in the pp-wave
 background, where there are two isometric directions associated to
 the two fibres in the Hopf decomposition of the transverse 3-spheres.
 
 The action (\ref{IIBaction}) is, again, a gauge fixed action, in which
 the $U(N)$ vector field, associated in this case to D3-branes wrapped
 on the two isometric directions, is set to
 zero. In this gauge $U(N)$ covariant derivatives reduce to ordinary
 derivatives, as in (\ref{covder2}).

  \subsection{The Matrix model}
 
 We can now particularize the action (\ref{IIBaction}) 
to the background (\ref{IIBppwave}). As in subsection 2.3 we take
light-cone gauge and
identify $x^+$ with the worldline time. We take as well the gravitons propagating
 in the $\psi$ direction, i.e. $k^\mu=\delta^\mu_\psi$.  Again, this is equivalent to taking
 the gravitons with momentum $p_-$, since $p_\psi$ and $p_-$ are related through the change
 of coordinates (\ref{coord}) as
 
 \begin{equation}
 p_\psi=-\mu L^2\,p_-\, .
 \end{equation}
 
 \noindent Therefore we describe the sector of the theory with light-cone momentum $p_-=-N/R$ taking
 $p_\psi=N$ and $R=\mu L^2$. Doing this we avoid the singularities that arise in the
 action if we simply take $k^\mu=\delta^\mu_-$, due to the fact that $k^2=g_{--}=0$.
 
In order to identify the second isometric direction we change coordinates

\begin{equation}
\xi=\frac{\chi+{\tilde \chi}}{2}\, ,\qquad
{\tilde \xi}=\frac{\chi-{\tilde \chi}}{2}
\end{equation}

\noindent and take $l^\mu=\delta^\mu_\xi$. This choice preserves the $\Z_2$ symmetry 
$\chi\leftrightarrow {\tilde \chi}$, $r\leftrightarrow y$, $x\leftrightarrow z$
of the background. Then we have

\begin{equation}
C^{(4)}_{\xi + ij}=\frac18 \mu r^4 \epsilon_{ijk} x^k\, ,\qquad
C^{(4)}_{\xi + ab}=\frac18 \mu y^4 \epsilon_{abc} z^c \, .
\end{equation}

\noindent These two potentials couple in both the CS and BI parts of the action. This will allow the existence of
zero energy solutions corresponding to expansions of the gravitons into the two 3-spheres
contained in the geometry.

We make the ansatz that the radii of the two 3-spheres are
commutative, consistently with the symmetries of the background, and
we restrict the use of capital letters for the non-commutative scalars.
We find for the CS action:

 \begin{equation}
 \label{CSIIB}
 S^{\rm CS}=\mu\int dx^+{\rm STr}\Bigl\{ \unity-\frac{1}{4L^2}(r^2+y^2)-i\frac{r^4}{16}
 \epsilon_{ijk}X^i X^j
 X^k-i\frac{y^4}{16}\epsilon_{abc}Z^aZ^bZ^c\Bigr\}\, ,
 \end{equation}

\noindent  and for the BI action
 
 \begin{eqnarray}
 S^{\rm BI}&=&-\mu\int dx^+{\rm STr}\Bigl\{\unity+\frac{r^2+y^2}{4L^2}-
 \frac{1}{2\mu^2L^2}\Bigl(\dot{r}^2+\dot{y}^2+\frac{r^2}{4}\dot{X}^2+\frac{y^2}{4}
 \dot{Z}^2\Bigr)+\nonumber\\
&&-\frac{1}{2\mu^2L^2}(\frac{r^2y^2}{r^2+y^2})\Bigl(\dot{\tilde \xi}-
\frac{A_i \dot{X}^i-{\tilde A}_a\dot{Z}^a}{2}\Bigr)\Bigl(\dot{\tilde \xi}-
\frac{A_j \dot{X}^j-{\tilde A}_b\dot{Z}^b}{2}\Bigr)+\nonumber\\
&&-\frac{1}{256}L^2 (r^2+y^2)\Bigl(r^4[X,X]^2
+y^4 [Z,Z]^2+2r^2 y^2 [X,Z]^2\Bigr)+\nonumber\\ 
&&+i\frac{r^4}{16} \epsilon_{ijk}X^iX^jX^k+i\frac{y^4}{16}\epsilon_{abc}Z^aZ^bZ^c\Bigr\}
 \end{eqnarray}

\noindent where we have expanded the action (\ref{IIBaction}) to leading order and
$A\equiv A_i dx^i$, ${\tilde A}\equiv {\tilde A}_a dz^a$.

Notice that ${\tilde \xi}$ only appears in the quadratic term in the second line of the
action. Integrating it out through its equation of motion we finally get

\begin{eqnarray}
\label{BI2IIB}
 S^{\rm BI}&=&-\mu\int dx^+{\rm STr}\Bigl\{\unity-
 \frac{1}{2\mu^2L^2}\Bigl(\dot{r}^2+\dot{y}^2+\frac{r^2}{4}\dot{X}^2+\frac{y^2}{4}
 \dot{Z}^2\Bigr)+\frac{r^2+y^2}{4L^2}+\nonumber\\
&&-\frac{1}{256}L^2 (r^2+y^2)\Bigl(r^4[X,X]^2+y^4 [Z,Z]^2
+2r^2 y^2 [X,Z]^2\Bigr)+\nonumber\\
&&+i\frac{r^4}{16} \epsilon_{ijk}X^iX^jX^k+i\frac{y^4}{16}\epsilon_{abc}Z^aZ^bZ^c
\Bigr\}
 \end{eqnarray}

 Combining (\ref{BI2IIB}) and (\ref{CSIIB}) and taking into account that $R=\mu L^2$ we 
 can finally read our proposal for the Type IIB Matrix model:

 \begin{eqnarray}
 \label{IIBmatrix}
 S&=&\int dx^+{\rm STr}\Bigl\{ \frac{1}{2R}\Bigl(\dot{r}^2+\dot{y}^2+
\frac{r^2}{4}\dot{X}^2+\frac{y^2}{4}\dot{Z}^2\Bigr)-
\frac{\mu^2}{2R}(r^2+y^2)+\nonumber\\
 &&+\frac{1}{256}R(r^2+y^2)\Bigl(r^4[X,X]^2+
y^4[Z,Z]^2+2r^2 y^2 [X,Z]^2\Bigr)+\nonumber\\
&&-i\frac{\mu}{8}r^4\epsilon_{ijk}
 X^iX^jX^k-i\frac{\mu}{8}y^4\epsilon_{abc}Z^aZ^bZ^c\Bigr\}
 \end{eqnarray}

\noindent Notice that this action is symmetric under the interchange $r\leftrightarrow y$, $X\leftrightarrow Z$,
like the background.

The Type IIB Matrix theory given by (\ref{IIBmatrix}) is a $U(N)$ gauge theory built up with six
non-Abelian scalars, $X^i$, $i=1,2,3$, and $Z^a$, $a=1,2,3$ plus two
Abelian ones, $r$ and $y$. The gauge field is set to zero through the gauge
fixing condition $A_\tau=0$.
In these coordinates the explicit symmetry
of the model is reduced to $(SO(3)\times U(1))^2$. Some comments about the
relation between our Type IIB Matrix model and the Tiny Graviton Matrix theory
of \cite{SJ} are now in order.

The Tiny Graviton Matrix theory of reference \cite{SJ} is a $U(N)$ gauge
theory, with  two main differences from our proposal.
First,  the Tiny Graviton Matrix theory is built up
with eight non-Abelian scalars plus an additional fixed
$U(N)$ matrix, ${\cal L}_5$, which is 
introduced in order to be able to
quantize the odd Nambu brackets of the light-cone
3-sphere. This matrix does not have a direct physical interpretation, but it allows
to couple
the RR 4-form
potentials in the action. Therefore, from our
point of view we would expect ${\cal L}_5$ to be
related to the existence of isometric directions in the background,
since in our construction the RR 4-form potentials couple in the action
contracted with the Killing vector associated to the isometry.
This is in agreement with the discussion in \cite{ST}. In this
reference it is argued that ${\cal L}_5$ has its origin in M-theory
compactified in $T^2$. In this compactification one of the directions  is the
light-cone direction, and the second one is the origin of ${\cal L}_5$
in the Type IIB theory. This idea becomes explicit in our
construction. Indeed, the two isometric directions of the Type IIB action are
the direction of propagation of the gravitons and the direction used
to construct this action from the action for type IIA gravitons using
T-duality. Therefore this action is related to an M-theory action with
two isometric directions. One is the direction of propagation of the
waves, in this case $x^-$, and the other one is the T-duality direction.

Second, the manifest symmetry of the Matrix
model in \cite{SJ} is the full $(SO(4))^2$ invariance of the
background. Our model has the advantage of not depending on the unphysical matrix ${\cal
  L}_5$ but at the expense of losing the full symmetry of the
background.

 The differences between the two models become more evident
when one looks at their vacuum solutions, as we do in the next subsection. We will see
however that both
models support fuzzy 3-sphere solutions with the right scaling
of the radius with the light-cone momentum in the large $N$ limit.

\subsection{The fuzzy sphere solutions}

A non-trivial check of the correctness of our Matrix model (\ref{IIBmatrix})
is that it supports
fuzzy 3-sphere solutions which agree exactly,
in the limit of large number of gravitons, with the classical 3-spheres
of \cite{GST,SS}. Note that the 3-sphere giant graviton expanding in the
spherical part of the geometry \cite{GST} and the one expanding in the $AdS$ part 
\cite{GMT,HHI} of the $AdS_5\times S^5$ spacetime are mapped under Penrose limit into the same
type of solution, a fact that is reflected in the action through the
$\Z_2$ symmetry 
$r\leftrightarrow y$, $X\leftrightarrow Z$. Therefore we only need to study in detail one of the two
solutions. 

Let us consider for instance
the dual giant graviton solution, i.e. the one expanding into the (Penrose limit
of the) $AdS$ part of the geometry. 

Our fuzzy 3-sphere ansatz is given by:

\begin{equation}
\label{ansatz1}
r={\rm constant}\, , \qquad y=Z^a=0\, , \, a=1,2,3\, , \qquad
X^i=\frac{1}{\sqrt{N^2-1}}J^i\, , \, i=1,2,3\, ,
\end{equation}

\noindent where $J^i$ are $SU(2)$ generators in an $N$ dimensional 
representation (in our conventions $[J^i,J^j]=2i\epsilon^{ijk}J^k$).
That is, we define the fuzzy
3-sphere as an $S^1$ bundle over a 
fuzzy 2-sphere 
. 
Substituting this ansatz in (\ref{IIBmatrix}) we get

\begin{equation}
\label{346}
S=\frac{N}{R}\int dx^+\, \frac{r^2}{2}\Bigl(\mu-\frac{r^2
  R}{4\sqrt{N^2-1}}\Bigr)^2\, .
\end{equation}

\noindent Since our configuration is static the Hamiltonian is just minus the Lagrangian, and we
can compare directly (\ref{346}) with the classical Hamiltonian of \cite{SS}, which is
given in our notation by \cite{LR}

\begin{equation}
H=-p_-\, \frac{r^2}{2}\Bigl(\mu+\frac{2\pi^2 T_3 r^2}{p_-}\Bigr)^2\, .
\end{equation}

\noindent We
find that both expressions agree exactly in the large $N$ limit, once we take into
account that $T_3=(8\pi^2)^{-1}$, in units in which $T_1=1$, 
and that we are describing the sector of the theory with
$p_-=-N/R$. The corresponding radii of the giant graviton solutions, given by

\begin{equation}
r^2=\frac{4\mu\sqrt{N^2-1}}{R}
\end{equation}

\noindent and 

\begin{equation}
r^2=-\frac{\mu p_-}{2\pi^2 T_3}\, ,
\end{equation}

\noindent also agree exactly in this limit. This is a non-trivial
check of the validity of our Matrix model (\ref{IIBmatrix}).

Note that in our construction the fuzzy 3-spheres are realized as $S^1$
bundles over fuzzy 2-spheres.
Therefore these vacua have just an explicit $U(1)\times SO(3)$
symmetry. Although at the classical level the $SO(4)$ covariance of the 3-sphere is
still present in a non-manifest way, the fuzzy 3-sphere consists on
an Abelian fibre over a non-Abelian base manifold, and this makes unclear how the whole
$SO(4)$ invariance can be recovered.
This set-up is however useful, because
the difficulties associated to the fuzzification of odd
dimensional spheres (see \cite{GR,Ram1,Ram2}) are avoided by reducing the dimensionality of the fuzzy
transverse space from 3 to 2, by means of writing the 3-sphere as an $S^1$ fibre over an
$S^2$ base manifold, and taking only the $S^2$ non-commutative.
This can be done consistently because the action that we use to describe the gravitons
contains an explicit Abelian Killing direction which can be identified
with the direction along the fibre\footnote{In this paper we have identified it with the
combination $(\chi+{\tilde \chi})/2$ in order to have a matrix model which is symmetric under
the interchange of the two $\R^4$ subspaces of the background.}.

On the other hand, 
the fuzzy 3-sphere vacua of the Tiny Graviton Matrix theory of
\cite{SJ} are fully $SO(4)$-symmetric\footnote{See also \cite{GR,Ram1,Ram2}.}. 
By adding ${\cal L}_5$ to the collection
of non-commutative transverse scalars it is possible to construct
finite dimensional representations of $SO(5)$ which can be further
reduced to $N$ dimensional representations of $SO(4)$. In this way the fuzzy
3-sphere is constructed from an intermediate fuzzy 4-sphere.
Comparing to our description,
this construction circumvents the difficulties associated to
the fuzzification of the 3-sphere by increasing by one the dimensionality
of the fuzzy space, which is precisely the role played by the matrix 
${\cal L}_5$.

Another difference between the fuzzy 3-sphere vacua of \cite{SJ} and the ones
constructed in this paper is that, although the solutions in \cite{SJ} have the correct
scaling of the radius with the momentum in the large $N$ limit, some non-commutativity
still remains
(see \cite{GR,Ram1,Ram2}). This is however not the case for the fuzzy
3-sphere solution constructed in this paper, which approaches neatly the classical 3-sphere
in the large $N$ limit, where all the non-commutativity disappears.

\section{Conclusions}

Using the action for coincident Type IIA gravitons constructed in \cite{JL1,JL2} \footnote{And
completed in this paper in order to describe the pp-wave background.} we have reproduced
Matrix String theory in the pp-wave background that is obtained by reducing the maximally
supersymmetric pp-wave background of M-theory  \cite{SY,DMS}.  We have also
clarified how in the Matrix model approach the fuzzy 2-sphere solutions of \cite{CS}, with 
non-vanishing $x^4$ position in the transverse space, emerge when  D0-brane
charge is induced in the configuration.

In the Type IIB case we have started
from the action describing coincident Type IIB gravitons of \cite{JLR}. This action
is connected by duality to the action for non-Abelian D-branes constructed in \cite{Myers}, and 
has been 
successfully used in the microscopical description of giant gravitons in $AdS_5\times S^5$
\cite{JLR}. Using this action
 we have made a proposal for a pp-wave Matrix model which supports fuzzy 3-sphere vacuum solutions with the right behavior in the large $N$ limit. 
 
Our Matrix model is a one dimensional gauge theory which could be a candidate for the
holographic description of strings in the pp-wave background. Indeed, 
$x^+$, which parametrizes the conformal boundary of the pp-wave \cite{BN}, is its
time direction, and the matrix model is compactified along $x^-$.

As we have mentioned there is a
second candidate for this holographic description\footnote{See also 
\cite{BKPS,SV,KKP,Plefka}.}, which is
the Tiny Graviton Matrix theory of \cite{SJ}.  We have already mentioned some differences between
the two constructions. On one hand
our Matrix
model does not depend on the matrix ${\cal L}_5$, 
which lacks a direct physical interpretation, however this happens at
the expense of losing the explicit
$SO(4)\times SO(4)$ symmetry of the transverse space, and of the Matrix model in \cite{SJ}.
This is related to the fact that
 ${\cal L}_5$ is associated to an isometry of
the background. Since we have made this isometry explicit in our construction we have reduced
the size of
the symmetry group.

 The existence of these two different Matrix models for the Type IIB pp-wave background
 could be related to the fact that there is no unique way to quantize diffeomorphisms in a 3-sphere. Therefore one could expect different gauge theories with the right continuum limit.
A possible connection between the two Matrix models, that would be interesting to check, is
 whether our Matrix model can be derived in the approach 
of \cite{SJ}
by first writing the 3-spheres in the transverse space as $U(1)$ fibres over $S^2$ base manifolds,
and then quantizing only the Nambu brackets associated to the three transverse scalars building
up each 2-sphere, which would now be even dimensional.

 Finally, we would like to mention that we have only constructed the bosonic
 parts of the Type II Matrix models, and that  it would be interesting to check
 their supersymmetry properties. Our starting point, the actions for Type
 II gravitons, are connected by dualities to (non-Abelian) D-brane actions, 
for which supersymmetry has been studied (see \cite{BBRS,Bilal}). Using dualities
 it should be possible to construct the 
fermionic parts of these actions. Moreover, since the pp-wave backgrounds
are linear perturbations to Minkowski we could use the results in 
\cite{TV}. In the Type IIB case we do not
 expect however that 
the invariance under the whole $PSU(2|2)\times PSU(2|2)\times U(1)$
 superalgebra of the pp-wave 
background \cite{SS2} will be manifest, due to the explicit breaking 
$SO(4)  \rightarrow SO(3)\times U(1)$ of our construction.

\subsection*{Acknowledgements}

We would like to thank B. Janssen and M.M. Sheikh-Jabbari for useful discussions. The work of Y.L. has been partially supported by CICYT grant BFM2003-00313 (Spain),
and by the European Commission FP6 program MRTN-CT-2004-005104, in
which she is associated to Universidad Aut\'onoma de Madrid.
D.R-G. would like to thank the Departamento de F\'{\i}sica Te\'orica y del Cosmos at Universidad de
Granada for its hospitality while part of this work was done and for partial financial support through
a visiting professor grant.


\end{document}